\documentclass[a4paper,11pt]{article}
\pdfoutput=1 

\usepackage{jinstpub} 
\usepackage{siunitx}
\usepackage{tikz}
\usepackage{pgfplots}
\usepackage{chemist}
\usepackage[utf8]{inputenc}
\usepackage[french]{babel}
\usepackage{caption}

\pgfplotsset{compat=newest}
\usepgfplotslibrary{external,colormaps} 
\title{High Rate RPC detector for LHC}

\author[i]{F. Lagarde \note{Corresponding author.}}
\author[a]{, A. Fagot} 
\author[a]{, M. Gul} 
\author[a]{, C. Roskas} 
\author[a]{, M. Tytgat}
\author[a]{, N. Zaganidis}
\author[b]{, S. Fonseca De Souza}
\author[b]{, A. Santoro}
\author[b]{, F. Torres Da Silva De Araujo}
\author[c]{, A. Aleksandrov}
% ...

\affiliation[a]{Ghent university, Dept. of Physics and Astronomy, Proeftuinstraat 86, B-9000 Ghent, Belgium}
\affiliation[b]{ Dep. de Fisica Nuclear e Altas Energias, Instituto de Fisica, Universidade do Estado do Rio de Janeiro, Rua Sao Francisco Xavier, 524, BR - Rio de Janeiro 20559-900, RJ, Brazil}
\affiliation[c]{Bulgarian Academy of Sciences, Inst. for Nucl. Res. and Nucl. Energy, Tzarigradsko shaussee Boulevard 72, BG-1784 Sofia, Bulgaria.}
% ...
\affiliation[i]{Universite de Lyon, Universite Claude Bernard Lyon 1, CNRS-IN2P3, Institut de Physique Nucleaire de Lyon, Villeurbanne, France.}
% ...

\emailAdd{f.lagarde@ipnl.in2p3.fr}

\abstract{The High Luminosity LHC (HL-LHC) phase is designed to increase by an order of magnitude the amount of data to be collected by the LHC experiments. The foreseen gradual increase of the instantaneous luminosity of up to more than twice its nominal value of \SI{10e34}{\per\centi\meter\square\per\second} during Phase I and Phase II  of the LHC running, presents special challenges for the experiments. The region with high pseudo rapidity ($\eta$) region of the forward muon spectrometer ($\num{2.4} >\left|\eta\right| > \num{1.9}$) is not equipped with RPC stations. The increase of the expected particles rate up to $\SI{2}{\kilo\hertz\per\square\centi\meter}$ ( including a safety factor 3 ) motivates the installation of RPC chambers to guarantee redundancy with the CSC chambers already present. The current CMS RPC technology cannot sustain the expected background level. A new generation of Glass-RPC (GRPC) using low-resistivity glass was proposed to equip the two most far away of the four high $\eta$ muon stations of CMS. In their single-gap version they can stand rates of few $\si{\kilo\hertz\per\square\centi\meter}$. Their time precision of about $\SI{1}{\nano\second}$ can allow to reduce the noise contribution leading to an improvement of the trigger rate. The proposed design for large size chambers is examined and some preliminary results obtained during beam tests at Gamma Irradiation Facility (GIF++) and Super Proton Synchrotron (SPS) at CERN are shown. They were performed to validate the capability of such detectors to support high irradiation environment with limited consequence on their efficiency.}

\keywords{Gaseous detectors; Resistive-plate chambers; Particle tracking detectors (Gaseous detectors); Materials for gaseous detectors}

\proceeding{XIV\textsuperscript{th} Workshop on Resistive Plate Chambers and related detectors\\
	19-23 February, 2018\\
	Puerto Vallarta, Mexico\\}

\begin{document}
\maketitle
\flushbottom
\section{Introduction}
In the actual CMS detector, all the muon stations are equipped with two kinds of muon detectors in order to ensure a good redundancy. In the barrel region, Drift Tubes (DT) and Resistive Plate Chambers (RPC) detectors are used, but Cathode Strip Chambers (CSC) and RPC for the end-caps region. So, RPCs cover all the regions except the high eta stations ($\left|\eta\right|> 1.6$) where only CSC detectors are present. To guarantee a redundancy in this region and improve the muon trigger efficiency it is planned to add new chambers during the long shut-down LS3. The projected increase of the LHC luminosity up to \SI{5e34}{\per\square\centi\meter\per\second} during the HL-LHC Phase 2 urges the need for new detectors with high rate capability \cite{a}.

\begin{figure}[htbp]
	\centering 
	\includegraphics[width=1\textwidth,clip]{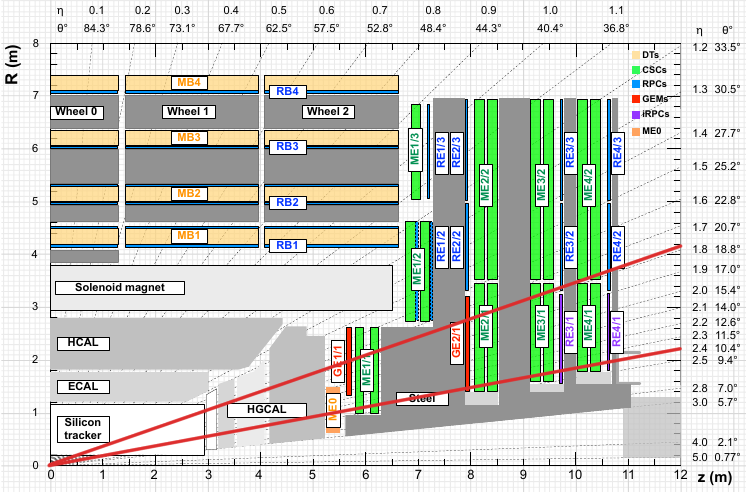}
	\caption{\label{fig:RPCUpgrade} An R-z cross section of a quadrant of the CMS detector, including Phase-1 and Phase-2  (RE3/1, RE4/1, GE1/1, GE2/1, ME0). The acronym iRPCs in the legend refers to the new improved RPC chambers RE3/1 and RE4/1. The interaction point is at the lower left corner. The locations of the various muon stations are shown in color (MB = DT = Drift Tubes, ME= CSC = Cathode Strip Chambers, RB and RE = RPC = Resistive Plate Chambers, GE and ME0	= GEM = Gas Electron Multiplier). M denotes Muon, B stands for Barrel and E for Endcap. The magnet yoke is represented by the darkest gray areas.}
\end{figure}

The instrumentation upgrade planned during the spectrometer upgrade project can be summarized in figure \ref{fig:RPCUpgrade}. 

It is proposed to equip the two first of the four high $\eta$ muon stations (GE1/1  \& GE2/1) by Gaseous Electron Multiplier (GEM). For the two last stations (RE3/1 \& RE4/1), RPC are proposed. In this region the expected rate during HL-LHC program was estimated with FLUKA \cite{c,d} and should not exceed $\SI{2}{\kilo \hertz \per \square \centi \meter}$, taking into account a safety factor 3 \cite{b}. Many RPC technologies were studied in order to equip these regions, of which the Doped Glass RPC \cite{4774543} technology was studied in order to install an improved  RPC in the higher $\eta$ region.

In its simple version, a generic RPC detector is made of two resistive plates whose outer face is covered by a resistive coating. The two plates play the role of the electrodes. The distance between the two electrodes is maintained constant using spacers. Between the two plates there is a constant gas mixture flow. As RPCs are gaseous detectors, their working principle is based on the ionization effect in the gas, caused by a charged particle passing through the detector volume. The avalanche multiplication is caused by a cluster of electrons obtained after the ionization. This leads to an electron charge developed inside the gap, which further drifts towards the anode.

The useful signal of the detector is the charge induced on the pickup electrode. The HV power supply is compensating the charge collected on the electrodes, by moving the charges outside the gap.

During this process the electric field is locally diminished and passage of other charged particles may go undetected. Once the charges are absorbed, the local electric field is then restored to its initial value. The resistivity of the electrodes used to eliminate possible sparks is also responsible of the limited RPC particle rate detection capability. By using low resistivity glass plates as electrodes (LRGRPC), the GRPC detectors particle rate capabilities can be increase. This new glass, developed by Tsinghua University, has a resistivity of the order of \SI{10e10}{\ohm\centi\meter} and a very high surface uniformity, with a roughness below \SI{10}{\nano\meter}\cite{4774543}\cite{WANG2010151}. Low resistivity is an important ingredient to reach high rate.

Previous results for this material can be found in reference \cite{Lagarde:2016fvf}. A chamber with the sizes close to the one required in the (RE3/1 \& RE4/1) was build using the Low Resistivity Glass. In this paper we will present the results obtained in beam tests at GIF++ with this chamber.

\section{Chamber description}
Due to producing method, the Tsinghua Low Resistivity Glass can not be produced larger than $\num{32}\times\SI{30}{ \square \centi \meter}$. The size limitation of the glass plates requires to find an efficient and robust way to assemble them in order to build large trapezoidal detectors for the CMS high $\left|\eta\right|$ region. In previous proceeding \cite{Lagarde:2016fvf}, two building methods by tessellation ("gluing fixation" and "mechanical fixation") have been studied in a cosmic stand. Both revealed to be equivalent in term of muon detection efficiency. 

\begin{figure}[htbp]
	\centering
	\includegraphics[width=0.6\textwidth,clip]{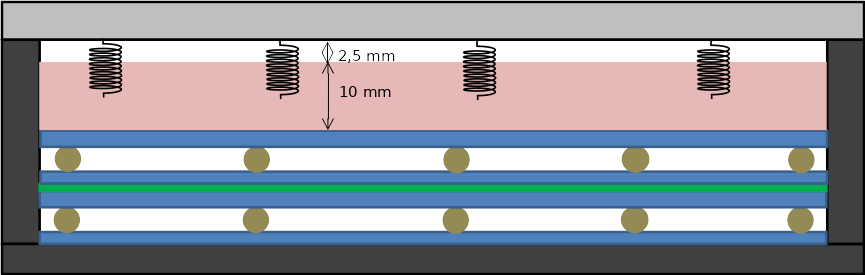}
	\caption{\label{fig:cassette} Scheme of the Cassette.}
\end{figure}

The second method has been selected to avoid the use of glue which could be radiation damaged. The glass tiles are maintained mechanically by the aluminium cassette using a PMMA plate, maintained in contact with the outermost electrode by springs compressing the detectors inside the cassette (figure.\ref{fig:cassette}).

\begin{figure}[htbp]
	\centering 
	\includegraphics[width=0.7\textwidth,clip]{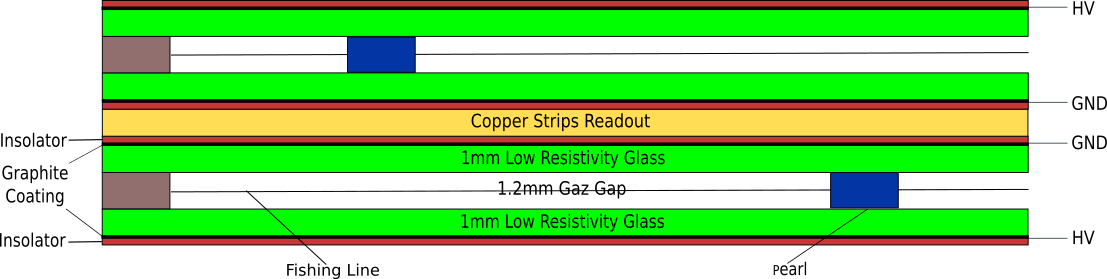}
	\caption{\label{fig:detector} Scheme of the Double Gaps detector.}
\end{figure}

The gas gap between the two electrodes is maintained fixed by pearled fishing lines (figure.\ref{fig:detector}). This design allows to gas tight the entire cassette rather than just the gas gap between the electrodes avoiding his inflating due to over-pressures. Very thin copper tapes are used to electrically connect the small glass together ( red squarres in figure \ref{fig:PCB}), forming an electrode. 

One large LRGRPC has been build this way (figure.\ref{fig:cassettepic}). The LR-GRPC configuration is the same as the actual instrumented RPC inside CMS (Double Gaps). A strip PCB similar to the CMS one is sandwiched between the two gaps. The CMS RPC's read-out electronics has been used.

\noindent
\begin{minipage}[th!]{.48\textwidth}
	\centering
\includegraphics[width=1\textwidth,clip]{NEWINT.png}
	\captionsetup{type=figure}\caption{\label{fig:cassettepic} Inner view of the cassette.}
\end{minipage}%
\hfill
\begin{minipage}[th!]{.48\textwidth}
	\centering
	\centering
\includegraphics[width=0.9\textwidth,clip,angle=180]{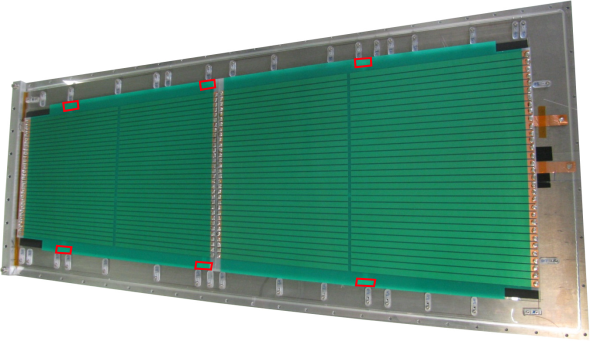}
	\captionsetup{type=figure}\caption{\label{fig:PCB} Inner view of the cassette with the top gap removed.}
\end{minipage}


\section{Results at the Gamma Irradiation Facility (GIF++)}
Beam tests were performed at the Gamma Irradiation Facility (GIF++) in order to characterize this detector.
\subsection{The Gamma Irradiation Facility (GIF++)}
\begin{figure}[htbp]
	\centering
	\includegraphics[width=0.8\textwidth,clip]{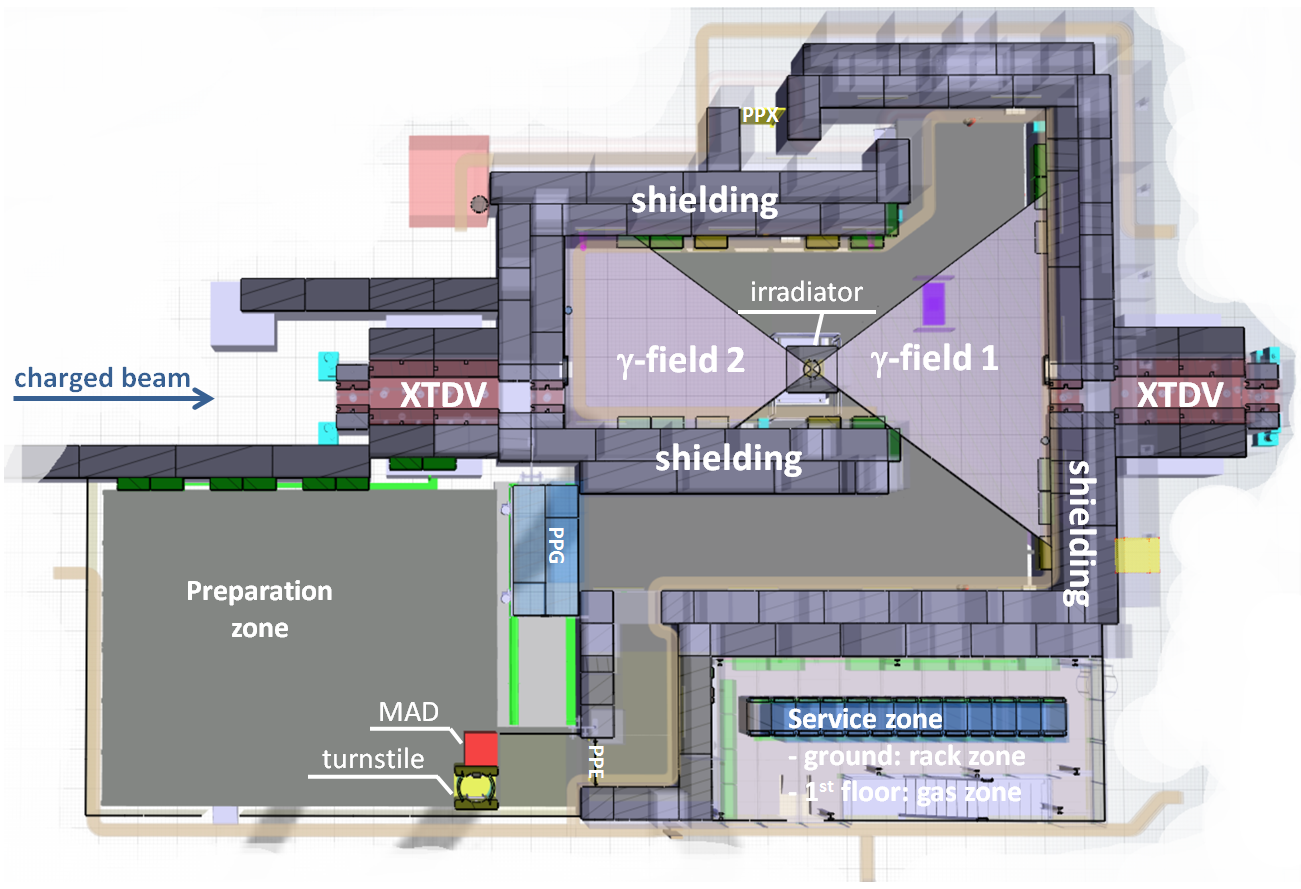}
	\caption{\label{fig:GIF} Scheme of the GIF++ bunker.}
\end{figure}

The Gamma Irradiation Facility (GIF++) is located at the H4 beam line at CERN. It provides a high energy charge particle beams (mainly muon beam with momentum up to 100 GeV/c) and a \SI{13}{\tera\becquerel} $^{137}$Cs radioactive source. This source produces a high background gamma environment (mainly $\gamma$ of \SI{662}{\kilo\electronvolt}) which allows to accumulate doses equivalent to HL-LHC experimental conditions in a reasonable time (figure.\ref{fig:GIF}). Two large radiation fields ($\pm$\SI{37}{\degree} horizontally and vertically) with individual radiation attenuation systems allow to vary the attenuation factor of the source from 1 to 46000. \cite{Pfeiffer:2016hnl}.

\subsection{Characterisation of the detector}

\begin{figure}[!ht]
	\centering
	\scalebox{1.3}{%
	\tikzsetnextfilename{TimeDistryBC}%
	\input{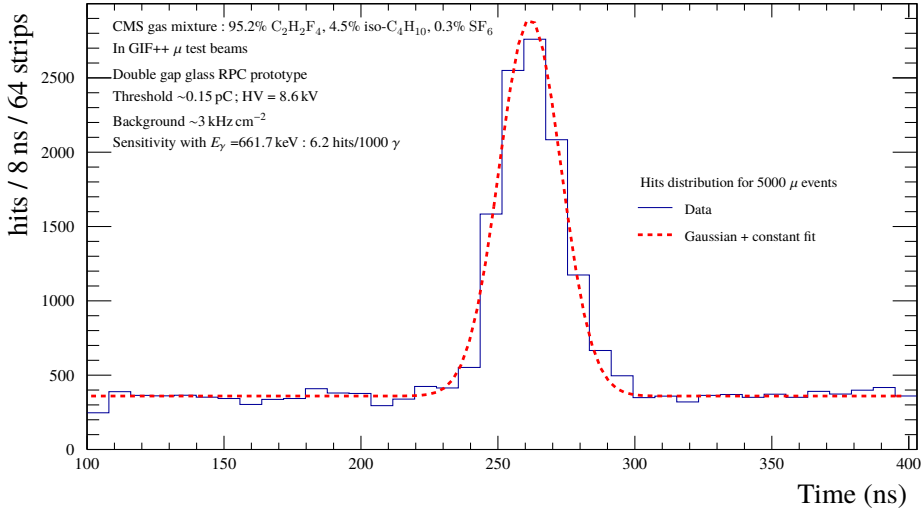}%
}
	\caption{Example of time distribution of the hits for \num{5000} events. Here, the signal is observed around $ t_0 =\SI{270}{\nano\second}$ and can be adjusted by the sum of a Gaussian of resolution $\sigma =\SI{11}{\nano\second}$ and a constant. The signal area corresponds to a window of $\pm 3 \sigma$ around $ t_0 $.}.
	\label{ajust}
\end{figure}

The strips are connected to a \SI{100}{\pico\second} TDC module writing continuously in a buffer memory. The coincidence of four Photo-Mutipliers placed before and after the chamber allow to detect the passage of charged particles and trigger the recording of a time window of the buffer TDC. The timestamp of the beginning of this selected window is realigned with \num {0} (figure \ref{ajust}).

In order to select the time window corresponding to the signal, the procedure bellow is performed :
Each timestamp of the hits of a strip is realigned with the average timestamp of the chamber by applying the formula:
\begin{equation}
T^{'}_{n} = T_ {n} -\bar{T} _ {strip\, n} + \bar{T}_{chamber}
\end{equation}
with $T^{'}_{n} $ the timestamp of the strip $n$ aligned, $T_{n}$ the timestamp of strip $n$ non-aligned, $\bar{T}_{strip \, n}$ the mean timestamp of the strip $n$ and $\bar{T}_{chamber}$ the mean timestamp of the chamber.
	
Then, a Gaussian plus constant fit of such distribution is performed using MINUIT\cite{310399}, and the time window corresponding to 
$T^{'}\in \left[mean-3\sigma,mean+3\sigma\right]$ is chosen as the signal time region.

The efficiency in the signal zone $\epsilon^{signal}$ is calculated as the number of triggers having at least one hit in the time zone region and in the spatial area affected by the beam.
\begin{equation}
\epsilon_{signal}=\frac{\sum\limits_{i=1}^{N^{trigger}} \delta^{trigger}_i(N^{hits}_{i})}{N^{trigger}} \mbox{where}
\left\{
\begin{array}{ll}
\delta^{trigger}_i(N^{hits}_{i})=1 & \mbox{if } N^{hits}_{i}>0 \\
\delta^{trigger}_i(N^{hits}_{i})=0 & \mbox{etherwise}
\end{array}
\right.
\end{equation}
where $N^{trigger}$ is the number of triggers.

In order to supress the efficiency due to noise inside the chamber the $\epsilon_{signal}$ must be corrected. To do so, the part of the time window before\footnote{The zone corresponding to the noise after the passage of the particles is not taken into account in order to avoid the increase of noise due to possible rebounds and crosstalk that could occur after their passage.} the beginning of the passage of the muons is cut off in intervals $I_{n}=\left[t_{n}-t_{win},t_{n}+t_{win}\right]$ where $t_{n}$ is the middle of the interval $n$ and $t_{win}$ the
half width of this interval. For each interval $n$, the average number of noise hits $\overline{N^{hits}_n}$ in the range $n$ is calculated:
\begin{equation}
\overline{N^{hits}_n}=\frac{\sum\limits_{i=1}^{N^{trigger}} N^{hits}_{n,i}}{N^{trigger}}
\end{equation}
with $N^{hits}_{n,i}$ the number of hits contained in the interval $n$ of the $i^{th}$ trigger.

The same procedure as for the signal zone is performed in order to obtain $\epsilon^{noise}_n$ corresponding to the efficiency that would have been found if the time zone was the $I_{n}$ interval.

The maximum rate of the number of hits is then selected $\overline{N^{hits}_{max}}$.

Finally, the corrected efficiency can be calculated using : $\epsilon=\frac{\epsilon^{signal}-k\epsilon^{bruit}}{1-k\epsilon^{bruit}}$
where,
\begin{equation}
k=\frac{1-\mbox{Poisson}\left(k=0,\lambda=\frac{\overline{N^{hits}_{max}}T^{signal}}{T^{noise}}\right)}{1-\mbox{Poisson}\left(k=0,\lambda=\frac{\overline{N^{hits}_{max}}}{T^{noise}}\right)}
\label{method2}
\end{equation}
supposing that the number of noise hits follow the Poisson law:
\begin{equation}
\mbox{Poisson}\left(k,\lambda\right)=\frac{\alpha^k}{k!}e^{-\alpha}. 
\end{equation} 
The factor $k$ weights $\epsilon^{bruit}$, in order to take into account the fact that the intervals $I$ and $T^{signal}$ are not equal. This factor represents the relationship between the probability to have at least one noise hit in the $T^{signal}$ interval and the probability that there is at least a noise hit in the interval $I$.

The efficiency value is taken as the center of the interval $\left[\epsilon, \epsilon^{signal}\right]$ where $\epsilon$ is the efficiency obtained by the formula \ref{method2} and $\epsilon^{signal}$ is the efficiency without correction. The total error on efficiency is given by:

\begin{equation}
\sqrt{{\frac{\left(\epsilon^{signal}-\epsilon\right)}{4}}^2+\epsilon_{stat}^2}
\end{equation}
where $\epsilon_{stat}$ is the statistical error.

\subsubsection{Efficiency and cluster size}
The detector was placed in the $\gamma$-field 2 region (figure\ref{fig:GIF}) at $\sim$\SI{3}{\meter} from the source. The sensibility to background of one gap RPC $r_{c}=0.31\%$ (the probability, that a \SI{662}{\kilo\electronvolt}$\gamma$ initiates a cluster in float glass RPC) is obtained with a GEANT4 based simulation of a standard Glass RPC. Thus, in the double gap scenario, the sensibility of the chamber to background is $2\times r_{c}=0.62\%$. 

\begin{figure}[h!]
	\centering
	\scalebox{1.}{%
	\tikzsetnextfilename{SigmoidBC}%
	\input{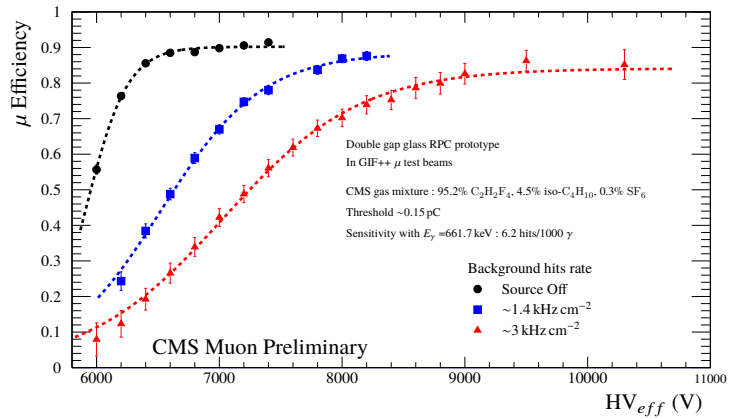}%
}
	\caption{\label{EFF}Muon detection efficiency with respect to the applied high voltage for different converted electrons rates.}
\end{figure}

\begin{figure}[h!]
	\centering
	\scalebox{1.}{%
	\tikzsetnextfilename{ClusterSizeBC}%
	\input{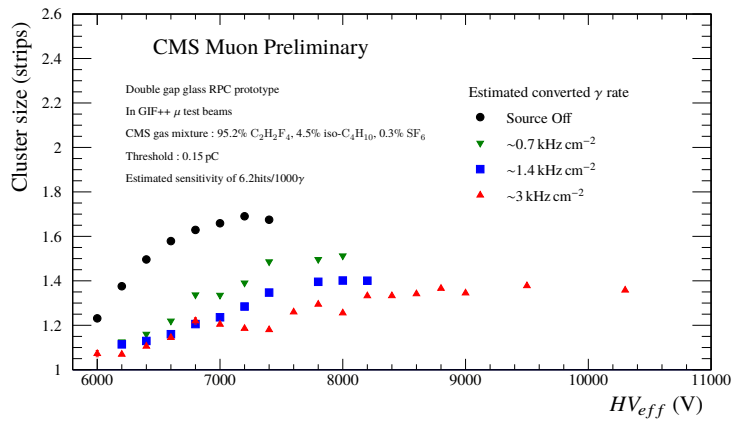}%
}
	\caption{\label{MUL}Cluster size with respect to the applied high voltage for different converted electrons rates.}
\end{figure}

\newpage
Taking into account the conversion factor and varying the attenuation factor it's possible to obtain the chamber efficiency (figure \ref{EFF}) and its cluster size (figure \ref{MUL}) with respect to the applied high voltage for different converted electrons rates.

The efficiency remains higher than \num{80}\% even for a converted electrons rate estimated to be \SI{3}{ \kilo \hertz \per \square \centi \meter}. The plateau moves more and more with respect to applied high voltages as the rate increases; this could be due to screen effects and the high dark current we observe for this chamber, thus decreasing the electric field applied between the two electrodes. The efficiency thus obtained however, is not optimal. Indeed, the center of the beam was close to the dead zone of the top gap, about \SI{2}{\centi\meter} between the two partitions, in the middle of the chamber. This dead zone is used to pass the cables collecting the signals of these partitions (figure \ref{fig:cassettepic}). Considering the fact that the beam was about \SI{2}{\centi\meter} wide, the loss of efficiency due to this dead zone has been estimated to be \num{5}\%.

For this chamber, the cluster size remains below \num{1.8} and decreases as the particle rate of the source increase. This reduction can be explained by the screen effect.


\section{conclusion}
Large double gap iRPC  using tessellation technique has been designed, assembled, and tested at GIF++, in order to study its characteristics. This prototype shows a good efficiency ($\sim\num{80}\%$) at \SI{3}{\kilo\hertz\per\square\centi\meter} despite the dead zone of one gaps and a reasonable cluster size. The start of the plateau was shown to vary of about \SI{2}{\kilo\volt} between no source and \SI{3}{\kilo\hertz\per\square\centi\meter}. This shift could be explained by the relatively high dark current observed during beam test. The test results show the acceptable parameters for the iRPC use at CMS and give the possibility for further improvements in the design and characteristics. 

\end{document}